\documentclass[authoryear,preprint,12pt]{elsarticle}

\usepackage{csquotes}
\usepackage{xcolor}

\usepackage{graphicx,subcaption}
\usepackage{amssymb}

\usepackage{lineno}
\usepackage{amsthm,amsmath,amsfonts}
\usepackage[a4paper, total={6.5in, 8.5in}]{geometry}

\biboptions{sort&compress}

\journal{Journal Name}

\begin{document}

\begin{frontmatter}


\title{Evaluating the effect of city lock-down on controlling COVID-19 propagation through deep learning and network science models}


\author{Xiaoqi Zhang\fnref{label2}}
\ead{xiaoqizh@buffalo.edu}
\author{Zheng Ji\fnref{label2}}
\ead{jz0429@163.com}
\fntext[label2]{National School of Development, Southeast University, Nanjing, China}

\author{Yanqiao Zheng\fnref{label3}}
\ead{zhengyanqiao@hotmail.com}
\fntext[label3]{School of Finance, Zhejiang University of Finance and Economics, Hangzhou, China}
\author{Xinyue Ye\fnref{label4}}
\ead{xinyue.ye@gmail.com}
\fntext[label4]{Urban Informatics-Spatial Computing Lab \& College of Computing, New Jersey Institute of Technoloy, New Jersey, U.S.A }

\author{Dong Li\fnref{label5}}
\ead{lidong@thupdi.com}
\fntext[label5]{Innovation Center for Technology, Beijing Tsinghua Tongheng Urban Planning \& Design Institute, China }

\begin{abstract}
The special epistemic characteristics of the COVID-19, such as the long incubation period and the infection through asymptomatic cases, put severe challenge to the containment of its outbreak. By the end of March, 2020, China has successfully controlled the within- spreading of COVID-19 at a high cost of locking down most of its major cities, including the epicenter, Wuhan. Since the low accuracy of outbreak data before the mid of Feb. 2020 forms a major technical concern on those studies based on statistic inference from the early outbreak. We apply the supervised learning techniques to identify and train NP-Net-SIR model which turns out robust under poor data quality condition. By the trained model parameters, we analyze the connection between population flow and the cross-regional infection connection strength, based on which a set of counterfactual analysis is carried out to study the necessity of lock-down and substitutability between lock-down and the other containment measures. Our findings support the existence of non-lock-down-typed measures that can reach the same containment consequence as the lock-down, and provide useful guideline for the design of a more flexible containment strategy.

\end{abstract}


\end{frontmatter}



\section{Introduction}
The novel coronavirus COVID-19 that was first reported in Wuhan, China at the end of 2019 quickly spread. Early 2020 has witnessed many efforts to contain the virus, such as the city lock-down, quarantining the suspected infectious cases and their close-contacts, setting health check point at crucial traffic nodes. By the mid of March 2020, the cumulative infectious cases have stopped growth in most of the major cities of China, including the epicenter of Wuhan. Although a growing list of published papers and reports claimed that the successful containment of COVID-19 in China was due to the national-wide travel ban and lock-down\citep{li2020,tian2020impact,qiu2020}, these studies focus exclusively on the aggregated number. The micro-mechanism how lock-down stops outbreak has rarely been analyzed based on real data. On the other hand, many countries, such as Italy, adopted similar lock-down policy, but failed to contain the outbreak of COVID-19 as China did. The South Korea and Japan didn't close up their major cities nor impose severe travel restriction to those uninfected people\citep{iwasaki2020,shaw2020,park2020}, but they both reported a low growth rate of infections within a relatively short time. Hence, it cannot be confirmed that the containment is achieved by lock-down unless the effect of other confounding non-lock-down measures, such as the conditional quarantine and social distancing, can be separated. 

To this end, a thorough investigation on the necessity of lock-down and its functioning mechanisms is needed. At the same time, due to the severe social-economic cost of lock-down \citep{barro2020,halder2011,kelso2020,bootsma2007,jorda2020,atkeson2020}, it is neither feasible for the other countries that are still struggling in the outbreak of COVID-19. Nor is an option for China given the risk of experiencing a second-wave outbreak. Therefore, alternative measures to the lock-down are recalled. A deep mechanism analysis of the lock-down can shed light on the searching of alternative containment measures and understanding their effectiveness, it is highly demanding at such a special moment when both US and China have experienced a second-wave outbreak of COVID-19 in recent weeks but neither of the top 2 economies can afford another round of lock-down.

In this paper, we attempt to explore the mechanism issue mentioned above. Our analysis is based on a novel network-based SIR (Susceptible-Infection-Recovery) model framework \citep{li2020,qiu2020} in which a time-lagged latent random infection mechanism is added to capture the epidemic characteristics of undocumented infectious and long incubation period, which cannot be handled in the classical SIR models \citep{keeling2011,heymann2020,li2020,mizumoto2020,wangliu2020,qiu2020,zou2020}. Unlike the model in \cite{li2020,qiu2020}, to capture hidden infection channels that are not directly linked to inter-regional population flow, such as the infection through panic-induced gathering\citep{wang2020,fang2020,garfin2020novel}, we won't adopt the prior assumption that the infection propagation cross regions can only be via the inter-regional population flow. Instead, a non-parametric network-based SIR model is applied, in which we do not impose any prior knowledge on the link weights across regions and let it be fully inferred from the COVID-19 outbreak data via deep learning methods. Using the inferred network, the connections between the infection link weights and the population flows are established through standard regression technique and tested for significance, by which, a counterfactual evaluation for the real effect of lock-down is carried out. Different from the counterfactual analysis done in existing literature \citep{qiu2020,li2020,tian2020impact,chinazzi2020} that attempt to justify the travel ban and the city lock-down measures as a sufficient condition for China's achievements in fighting COVID-19, we evaluate the necessity of these measures in the sense whether or not there exists much more moderate prevention measures that are as effective as the travel ban and city lock-down in terms of containing the outbreak of COVID-19, while have less negative impact on the social-economic development. We give positive evidence for the existence of such alternative measures, and also discuss the substitutability between lock-down and the other containment measures. We highlight that the substitutability can help quantitatively design the combination of containment measures that reach the balance between containing COVID-19 and the social-economic cost.
\section{Methodology}
\subsection{Model construction}
The NP-Net-SIR model is set up as the following:
\begin{equation}\label{network sir}
\begin{aligned}
\frac{d\mathbf{n}(t)}{dt}&=\int_{t-incub}^{t}p_{I}(\tau,t) \mathbf{W}(t) \cdot\mathbf{n}(\tau)d\tau-r(t)\mathbf{n}(t)\\
\mathbf{m}(t)&=\int_{t-incub}^{t}p_{B}(\tau,t) \mathbf{n}(\tau)d\tau
\end{aligned}
\end{equation} 
where $\mathbf{n}(t)=(n_{1}(t),\dots,n_{k}(t))^\top$ is the vector of the cumulative number of infectious cases of $k$ regions by time $t$, both the documented and undocumented cases are included within $\mathbf{n}(t)$, $\mathbf{m}(t)=(m_{1}(t),\dots,m_{k}(t))^\top$ denotes the documented number of infectious case by time $t$, $r(t)$ is the time dependent recovery rate. To reflect the epidemic characteristics of COVID-19 that the incubation period (14 days) is very long \citep{heymann2020,li2020,mizumoto2020,wangliu2020,qiu2020,zou2020} and the asymptomatic infectious cases can proceed the transmission, we add two time-dependent probability $p_{B}(\cdot,t)$ and $p_I(\cdot,t)$, they capture the time-lagged randomness within the two processes that the hidden infectious cases get discovered ($p_B$) and that the hidden infectious cases transmit the virus to healthy people ($p_I$). Without loss of generality, we let $r$, $p_B$ and $p_I$ depend on time continuously so as to capture the impact of time and the government prevention measures.

To formulate the spatial interactions of COVID-19 outbreak, we set a family weighted network adjacency matrices $\{\mathbf{W}(t)\}$ with $\mathbf{W}_{ij}(t)\in [0,1]$ for all $ij$ entries and all $t$. The adjacency matrix $\mathbf{W}(t)$ captures the cross-regional link weight of COVID-19 outbreak and can be interpreted as the proportion of the past cumulative infectious cases in region $j$ that contribute to the newly infected cases in region $i$. In previous studies \citep{li2020,qiu2020}, the adjacency matrices are directly identified as a constant multiple of the population flow matrix cross regions. This assumption is not sufficient to capture outbreak channels other than the point-to-point travel, such as the infection by panic-driven gathering, multi-destination travelling and the like \citep{wang2020,cohen2020,ferguson2020,pueyo2020,harris2020}. To account for these hidden channels, we take the non-parametric specification of $\mathbf{W}$ rather than impose prior knowledge. We also let $\mathbf{W}$ continuously depends on time $t$, accounting for the effect of time and various prevention measures. Model \eqref{network sir} is trained by deep learning technique, and the details are presented in \ref{training}.

\subsection{Counterfactual evaluation on the effects of travel ban and lock-down}\label{counter}
The effects of travel ban and city lock-down on containing the outbreak of COVID-19 can be evaluated based on the population flow data from Baidu Migration Index (available through the url \enquote{https://qianxi.baidu.com/}) and the trained NP-Net-SIR. The temporal network adjacency matrix $\mathbf{W}(t)$s sketches the cross-regional outbreak link strength of COVID-19 and its variation trend over time. The variation of $\mathbf{W}(t)$ is by and large the consequence of the travel restriction policies, but as we comment in the introduction, it cannot exclude the impact of panic and the other type of unaware driving force. To single out the impact of travel ban and city lock-down, we apply the following regression analysis:
\begin{equation}\label{reg}
\mathbf{W}_{kj}(t_i)=\alpha+\beta \overline{\mathbf{T}}_{kj}(t_i)+\varepsilon_{kji}
\end{equation}
where the temporal matrix $\overline{\mathbf{T}}(t_i)$ is the weighted average of the singe-day population flow matrices $\mathbf{T}(t_i)$s (extracted from Baidu migration index) by the infection probability $p_I(\cdot,t_i)$
\begin{equation}\label{wsT}
\overline{\mathbf{T}}(t_i)=\sum_{j=0}^{incub-1}p_I(j,t_i)\mathbf{T}(t_i-incub+j).
\end{equation}

Since Wuhan City was locked down, a bucket of containment measures had been applied by both Wuhan and the other major cities in China, such as the social distancing, conditional quarantine and setting health checkpoint in major transportation facilities etc., all of which could affect the link weights and contribute to contain the outbreak of COVID-19. To differentiate the effect of lock-down from the other measures, we fit the equation \eqref{reg} only using the data with Jan. 23, 2020 when Wuhan start to lock down. Given the estimates coefficient $\hat{\alpha}$, $\hat{\beta}$, the estimate to the residuals $\hat{\varepsilon}_{kji}$ for the $t_i$s after Jan. 23, 2020 are calculated and interpreted as the part of infection link weights unexplainable by population flows, accounting for the effect of the non-lock-down measures. Fix $\hat{\alpha}$, $\hat{\beta}$ and $\hat{\varepsilon}_{kji}$s, \eqref{reg} will be used to evaluate the impact of counter-factually increasing the population flow intensity between region pairs.

Unlike the existing studies \citep{tian2020impact,li2020,qiu2020,fang2020,chinazzi2020,zhang2020,anderson2019} that focus almost exclusively on the sufficiency question, i.e. whether lock-down really help mitigate the outbreak of COVID-19, this study attempts to answer the inverse problem. That is the necessity of lock-down, i.e. whether or not there exists an alternative prevention strategy that causes less damage to the social-economic development while performs as effective as the travel ban and lock-down in containing the outbreak of COVID-2019. Since the existence of such alternatives might be timing-sensitive, we consider different initialization time $t^s$ for the counterfactual worlds in which the travel ban and city lock-down are relaxed. The degree of relaxation is measured by a proportion $r_{jk}$ for each pair of destinations $j$ and $k$ such that after the the relaxation, the traffic flow intensity is increased to $\mathbf{T}^r_{jk}(t_i)=(1+r_{jk})\times \mathbf{T}_{jk}(t_i)$ for $t_i\geq t^s$. Given the relaxed population/traffic flow matrix $\mathbf{T}^r(t_i)$s, we can update the adjacency matrix $\mathbf{W}(t_i)$ to $\mathbf{W}^r(t_i)$ via \eqref{reg}. Denote $\mathbf{r}$ as the matrix consisting of all $r_{jk}$s, then we evaluate the necessity of travel ban by asking whether there exist a positive $\mathbf{r}$ matrix (with at least one positive entry and no negative entries at the mean time) such that under the updated $\mathbf{W}^r(t_i)$s by $\mathbf{r}$, the outbreak status of COVID-19 are no worse than the current for every $t_i\geq t^s$. The comparison of outbreak status between the real case and the counterfactual case can be measured in various different ways. In this study, we focus on three measures that are summarized through the following three constraints:
\begin{equation}\label{R0 compare}
R_0(\mathbf{W}^r(t_i),r(t_i))\leq \max(R_0(\mathbf{W}(t_i),r(t_i)),1),\,\forall t_i\geq t^s
\end{equation}
\begin{equation}\label{infect compare}
\mathbf{m}^r(t_i)\leq \mathbf{m}(t_i),\,\forall t_i\geq t^s
\end{equation}
\begin{equation}\label{death compare}
\mathbf{D}^r(t_i)\leq \mathbf{D}(t_i),\,\forall t_i\geq t^s
\end{equation}
where $R_0$ is the basic reproduction number which depends on the maximal eigenvalue of adjacency matrix and recovery rate. $\mathbf{m}^r(t_i)$ denotes the estimated documented infectious case by the updated $\mathbf{W}^r$ matrix through \eqref{network sir}. $\mathbf{D}(t_i)$ ($\mathbf{D}^r(t_i)$) is the total death cases (updated by $\mathbf{r}$) by time $t_i$ which depends on both the total number of infections and the local healthcare resources, the detailed calculation of $\mathbf{D}(t_i)$ is presented in \ref{death calc}.

These three constraints refer to three different goals of prevention, which require that after relaxation, the total number of infection and death shouldn't be greater than their current value, $R^r_0$ shouldn't induce infection divergence (greater than 1) or at least shouldn't be greater than its current value. The \enquote{no greater than} relation in \eqref{R0 compare}-\eqref{death compare} is in the point-wise sense, i.e. it has to hold for all region and all time after $t^s$, therefore, it is a very stringent restriction on the relaxation. Formally, any non-trivial relaxation $\mathbf{r}$ matrix satisfying the constraints corresponds to a Pareto improvement of the current prevention strategy. In our counterfactual analysis, we shall search for each constraint type the Pareto optimal relaxation strategy $\mathbf{r}^\ast$ from which no further Pareto improvement is allowed. This Pareto optimal $\mathbf{r}^\ast$ has practical significance in guiding the containment measure design for those countries suffering from COVID-19 now.
\subsection{Counter-factual evaluation on the substitutability between lock-down and other non-lock-down measures}\label{counter alter}
Except for city lock-down, many other non-lock-down measures have also been utilized to prevent COVID-19 outbreak, such as the \enquote{social distancing} \citep{pike2020,zhang2020}. All these measures can contain the outbreak of COVID-19, while compared to travel-ban and city lock-down, they generate less negative impact on the social-economic development, meanwhile their application is more accurately targeted rather than applies for all people regardless their healthiness and vulnerability to COVID-19. It can be reasonably expected that the execution of these non-lock-down measures can by and large substitute the lock-downs and reduce the harm to the economy induced by lock-down. 

To quantify the substitutability, we extend \eqref{reg} to include the effect of non-lock-down measures. We roughly divide all the non-lock-down measures to two classes, which are the measures adopted by the flow-in regions and the measures by the flow-out regions. The flow-in measures, its effect is quantified as a parameter $in_k(t_i,t_a)$, include the quarantine of arriving travellers from out-town, the close up of schools, the cancelation of gathering public activities and so on; the flow-out measures, its effect quantified as another parameter $out_j(t_i,t_b)$, include setting health check point in the entrance of inter-regional high-way, airport, rail stations and so on. The notation $k$, $j$ represent the index for regions, $t_a$ and $t_b$ represent the starting date of the relevant measures, in the other words, we let 
\begin{equation}
\begin{aligned}
in_k(t_i,t_a)=&\begin{cases}
\gamma_k,&t_i\geq t_a\\
0,&else\end{cases}\\
out_j(t_i,t_a)=&\begin{cases}
\theta_j,&t_i\geq t_b\\
0,&else\end{cases}
\end{aligned}
\end{equation}
for some constant magnitude parameter $\gamma_k$s and $\theta_j$s that measure the execution strength of relevant measures. Adding $in_k(t_i,t_a)$ and $out_j(t_i,t_b)$ into \eqref{reg} yields the following regression:
\begin{equation}\label{RD}
W_{kj}(t_i)=\alpha+(\beta-out_{j}(t_i,t_b))\cdot \overline{\mathbf{T}}_{kj}(t_i)-in_{k}(t_i,t_a)+\varepsilon'_{kji}
\end{equation}
where we suppose the flow-in measures impact the infection adjacency matrix additively while the flow-out measures impact through a multiplier of the population flow. For the starting date of two classes of non-lock-down measures, we follow the timeline provided in \cite{tian2020impact} and set $t_a$ as Jan. 26, 2020 when all 31 provinces in mainland China had already initiated the first-class protocol for emergent public health event which include the execution of various quarantine measures and the close-up of major public facilities. $t_b$ is set to Jan. 30, 2020 when health check point had been set at all major high-way entrances, railway stations and airports within Mainland China. 

 Given the estimate to parameter $\gamma_k$s, $\theta_j$s and the residuals $\varepsilon'_{jki}$, the counterfactual analysis is done by solving the same set of Pareto optimization problem under the same constraints as in the previous section. The only difference is that in the current setting, not only the relaxation matrix $\mathbf{r}$, but the set of non-lock-down parameter $\gamma_k$s, $\theta_j$s can also be simultaneously adjusted. 

\section{Results}
\subsection{Goodness of model fitting}
The NP-Net-SIR model is trained by using the province-level daily infection data collecting during Jan. 10 - Mar. 8, 2020 and from the official website of the National Health Commission (NHC) of China.

\begin{figure}[t!]

        \includegraphics[height=8cm,width=15cm]{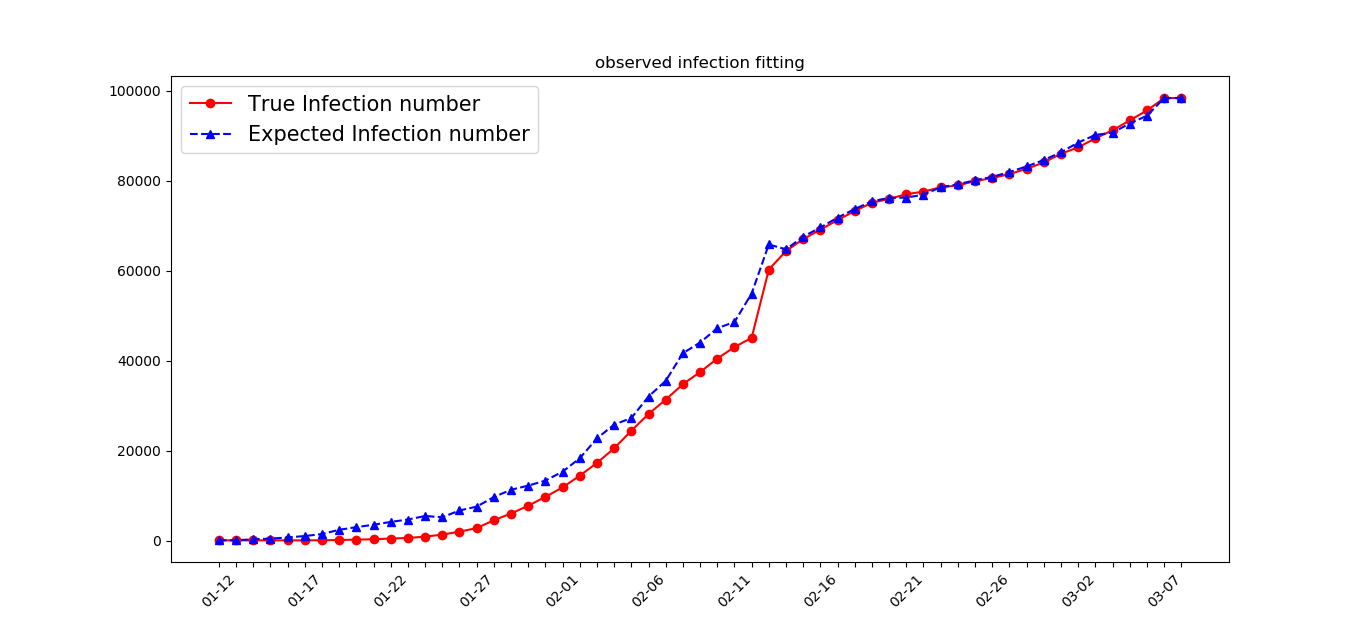}

     \caption{Aggregated documented infectious cases over all 31 provinces in Mainland China}\label{goodness of fitting}
\end{figure}

\begin{figure}[t!]

        \includegraphics[height=12cm,width=17cm]{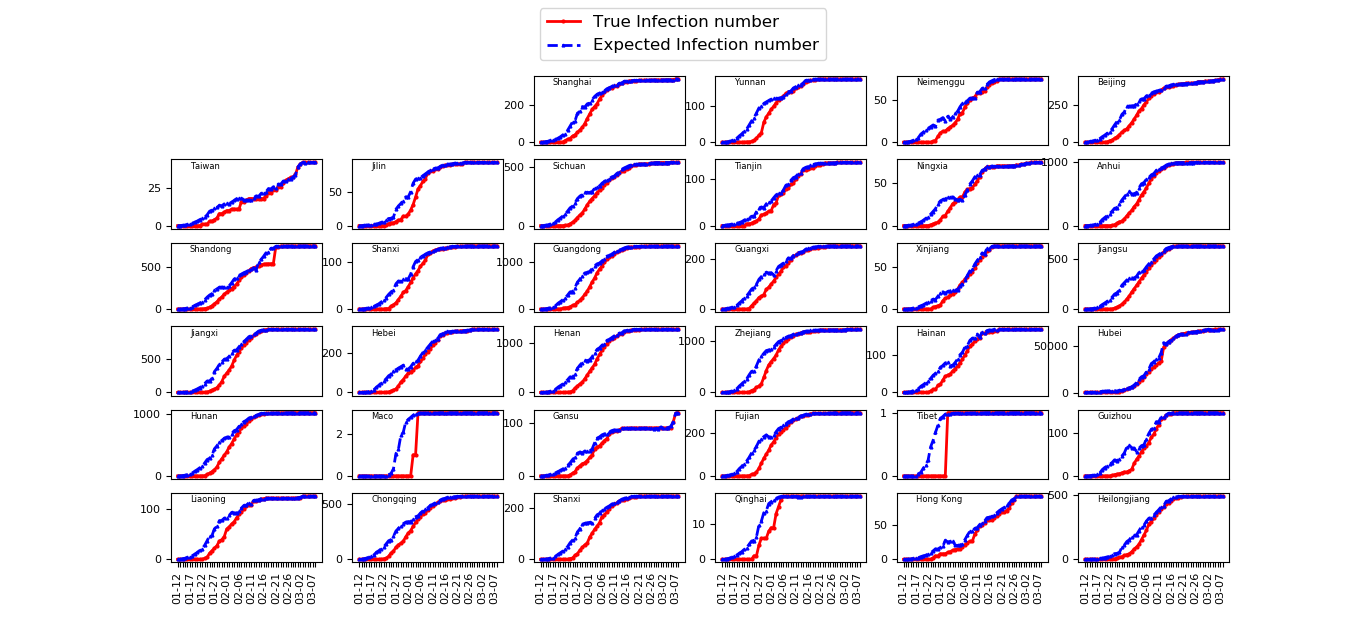}

     \caption{Documented infectious cases for 31 provinces in Mainland China}\label{goodness of fitting1}
\end{figure}

Fig. \ref{goodness of fitting} and \ref{goodness of fitting1} present respectively the fitting to temporal variation trend of documented infectious case from Jan. 10, 2020 to Mar. 7, 2020 for the national-wide aggregation case and province-level case for all 31 provinces in mainland China. Table \ref{table: accuracy} reports the fitting accuracy $R^2:=\frac{\Vert\hat{\mathbf{m}}-\mathbf{m}\Vert^2}{\Vert\mathbf{m}\Vert^2}$ measuring the relative difference between the estimated ($\hat{\mathbf{m}}$) and the real ($\mathbf{m}$) documented infection number since Feb. 12, 2020. It is quite apparent that the fitting accuracy after the Feb. 12, 2020 for all situations in the two figures are extremely high ($R^2>0.99$ for the aggregation over the whole China), and the fitted number is systematically greater than the reported number before Feb. 12, 2020. This is due to that we set Feb. 12, 2020 as the change point before which we do not punish the positive estimation error so as to reflect the potential under-estimate of the report data. The high accuracy after Feb. 12, 2020 demonstrates the explanation power of our NP-Net-SIR model. As a comparison, we run the classic SEIR model with the version discussed in \cite{li2020} on the same data set, and calculate the $R^2$ measure for both model after Feb. 12, 2020, the result shows our model performs much better by lifting $R^2$ by 12\%. The difference between the \enquote{over-estimated} infectious cases by our model and the reported cases before Feb. 12, 2020 can be thought of as a measure to the hidden infectious case that are not counted in the statistics. We calculate the ratio of the hidden cases and the total cases, finding that on the national-wide level, there were 79.27\% of hidden cases on average that were not reported before Feb. 12, 2020, this ratio is close to the one reported in \cite{tian2020impact}. If we look at the province-level data, the hidden ratios exceeds 90\% for most of the provinces in mainland China, among which Fujian, Guizhou, Yunan, Jiangsu, Jiangxi and Shanxi provinces are the top 6 with hidden ratios greater than 96\%, while Hubei is the province with lowest hidden ratio (70\%). This outstanding hidden ratio of Hubei can be attributed to the fact that Hubei is the epicenter of the COVID-19 outbreak within China, which was attacked by COVID-19 in the earliest time, and also reacted earliest in time to the virus. In contrast, all the other provinces suffered from the transmit-in cases in the early stage and therefore failed to react in time and cause a significant delay of updating the number. 

\begin{minipage}{\linewidth}
\begin{center} 
\captionof{table}{Estimation Accuracy by $R^2$}\label{table: accuracy}
\centering
\resizebox{3cm}{7cm}{%
\begin{tabular}{ll}
\hline\hline
Province & $R^2$\tabularnewline
\hline
Full country & 0.996\tabularnewline
Shanghai &0.979\tabularnewline
Yunnan &0.967\tabularnewline
Neimenggu & 0.979\tabularnewline
Beijing &0.972\tabularnewline
Taiwan &0.978\tabularnewline
Jilin &0.977\tabularnewline
Sichuan & 0.975\tabularnewline
Tianjin & 0.991\tabularnewline
Ningxia & 0.979\tabularnewline
Anhui &0.968\tabularnewline
Shandong & 0.979\tabularnewline
Shanxi &0.979\tabularnewline
Guangdong &0.971\tabularnewline
Guangxi &0.974\tabularnewline
Xinjiang &0.993\tabularnewline
Jiangsu &0.976\tabularnewline
Jiangxi &0.977\tabularnewline
Hebei &0.976\tabularnewline
Henan &0.976\tabularnewline
Zhejiang &0.965\tabularnewline
Hainan &0.981\tabularnewline
Hubei &0.997\tabularnewline
Hunan &0.974\tabularnewline
Maco &0.897\tabularnewline
Gansu &0.976\tabularnewline
Fujian & 0.964\tabularnewline
Tibet &0.907\tabularnewline
Guizhou &0.967\tabularnewline
Liaoning &0.973\tabularnewline
Chongqing &0.978\tabularnewline
Shanxi &0.968\tabularnewline
Qinghai &0.963\tabularnewline
Hong Kong &0.984\tabularnewline
Heilongjiang &0.984\tabularnewline
\hline\hline
\end{tabular}%
}
\end{center}
\end{minipage}\\

\subsection{Overview of the infection dynamics}
Due to the close connection between cross-regional population flow and the inter-regional outbreak of COVID-19 claimed in the literature \citep{qiu2020,li2020}, we present an overview on the strength of this connection in the following Fig. \ref{qq_wuhan} and \ref{co-variation}, in which the correlation between the (estimated) temporal infection adjacency matrix $\mathbf{W}(t)$ and the traffic flow matrix $\mathbf{\mathbf{T}}(t)$ are visualized in different manners. The first line subplots of Fig. \ref{qq_wuhan} consists of the scatter plots of all $\mathbf{\overline{\mathbf{T}}}_{kj}(t_i)$s versus $\mathbf{W}_{kj}(t_i)$s before (left) and after (right) the time of Wuhan lock-down (Jan. 23, 2020). The second and third lines of Fig. \ref{qq_wuhan} plot only those $\mathbf{\overline{\mathbf{T}}}_{kj}(t_i)$ and $\mathbf{W}_{kj}(t_i)$s that are end up with (the second line) or sourced from (the third line) Hubei province. To make the variation trend of the relation between $\mathbf{W}(t)$ and $\mathbf{T}(t)$ clearly visible,  we only plot the entries of $\mathbf{W}(t)$ and $\mathbf{T}(t)$ for five dates before and after Jan. 23, 2020. Fig. \ref{co-variation} gives the temporal view of the variation trend of the total flow-in (first line) and flow-out (second line) infection link weight and traffic flow intensity since Jan. 19, 2020 and for the top 7 provinces (we rank all 31 provinces by their aggregated flow-in and flow-out infection link weight averaged up to Jan. 23, 2020 in the descending order, and plot the data for first 7 provinces in each of the flow-in and flow-out category). In all the plots, we make the log transform for entries in $\mathbf{W}(t)$ and $\mathbf{T}(t)$, the horizontal axis corresponds to the entries of $\mathbf{T}(t)$ vertical axis corresponds to $\mathbf{W}(t)$. For the entries of $\mathbf{W}(t)$, we rescale it first by the potential infection number, i.e. $\mathbf{W}_{kj}(t)\cdot \frac{\mathbf{n}_k(t)}{\mathbf{n}_j(t)}$ before taking log transform. By rescaling, we hope to take the effect of the stock number of potential infections into account.

From the first line of Fig. \ref{qq_wuhan}, a counter-intuitive finding is that the correlation between population flow intensity across regions and the infection link weight is quite weak, no matter before or after Jan. 23, 2020. Especially in Fig. \ref{qq_wuhan}, a great portion of scatter points are clustered around a straight line close and parallel to the horizontal axis, such an observation does not support a linear correlation exists between $\mathbf{W}(t)$ and $\mathbf{T}(t)$ as imposed in \cite{li2020,qiu2020}. The the subfigure \ref{qq_wuhan}c. and \ref{qq_wuhan}e. do show a significant linear correlation between population flow strength and the infection link weight at least before the lock-down, while the flow-out population before Jan. 23, 2020 turns out more powerful in spreading the virus as in the subfigure \ref{qq_wuhan}e., the scale of the vertical axis is much greater than that in the subfigure \ref{qq_wuhan}c.. But on the other hand, after Jan. 23, 2020, the linear relationship between $\mathbf{W}(t)$ and $\mathbf{T}(t)$ gets sharply decayed, after Feb. 10, 2020, the correlation coefficient between entries of them cannot be differentiated from $0$ no matter for either the flow-in or the flow-out population. This observation contradicts to the classic assumption in \cite{li2020,qiu2020}. In fact, by the linear correlation assumption, the city lock-down can only control the number of people moving across regions, it has nothing to do with the proportion of infectious cases within these migrants, which should be a constant if only the lock-down and/or travel ban measures are applied. In the other words, if lock-down really works to contain the outbreak, the scatters in the right panel of Fig. \ref{qq_wuhan} should converge gradually to the origin along with a straight line with positive slope, rather than all scatters rotated toward the horizontal axis as shown in Fig. \ref{qq_wuhan}. The finding implies that the measures that really help contain the outbreak of COVID-19 may not be the lock-down, instead, they should be the other measures initiated almost simultaneously with the lock-down and their effect confounded with that of lock-downs. To correctly evaluate the real effect of each type of containment measures, we have to differentiate the confounding measures and their impact, which we shall leave to the discussion in section \ref{alter measures}.

By Fig. \ref{co-variation}, there exists an significant gap period around one week between the vanishing of the flow-in and -out infection link weight and the decay of the corresponding population flow intensity. For the time series of flow-in and -out population intensity, all top 7 provinces reached their minimum before Jan. 31, 2020, while at the mean time none of them have made the flow-in and -out infection link weight decayed to somewhere close to zero until Feb. 6, 2020. Such an one week gap period reflects the effect of the long incubation period of the COVID-19 and the fact that its infection can happen via infectious cases without symptoms. The classical SIR/SEIR models ignore this gap period and tend to over-estimate the basic reproductive number $R_0$ in the early stage which would trigger the most severe containment measures, such as the lock-down, if the decision is made upon that base. 

In sum, from the brief overlook on the numerical relationship between infection link weight $\mathbf{W}$ and the population flow intensity $\mathbf{T}$, we can summarize: 1) the positive linear correlation assumption made in many versions of the SIR/SEIR model \citep{tian2020impact,li2020,qiu2020,fang2020,efimov2020} does not hold uniformly during the outbreak of COVID-19, but it does hold for the population flow into and out of Hubei province before the great lock-down; 2) after the lock-down initiated since Jan. 23, 2020, the positive linear correlation between $\mathbf{W}$ and $\mathbf{T}$ is reduced significantly and fastly to zero and this reduction shouldn't be simply attributed to the contribution of lock-down, the effect of other confounding measures should be examined more carefully; 3) an one-week gap-period exists between the decay of population flow intensity and infection link weight, which should be a consequence led by the epidemic characteristics of COVID-19, the classical SIR/SEIR model neglects this gap-period and can lead to too severe containment measures.

\begin{figure}[t!]

        \includegraphics[height=10cm,width=16cm]{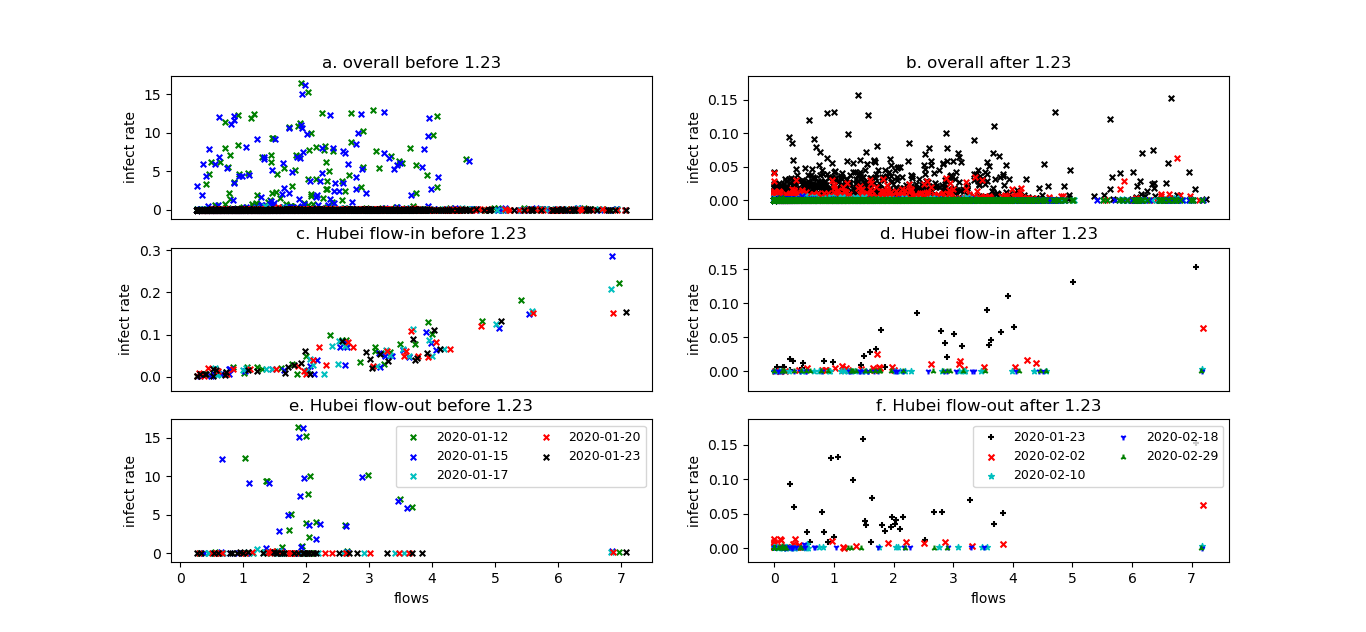}

     \caption{Relationship between infection link weight $\mathbf{W}$ and population flow intensity $\mathbf{T}$ with Hubei as origin/destination province}\label{qq_wuhan}
\end{figure}
\begin{figure}[t!]

        \includegraphics[height=10cm,width=15cm]{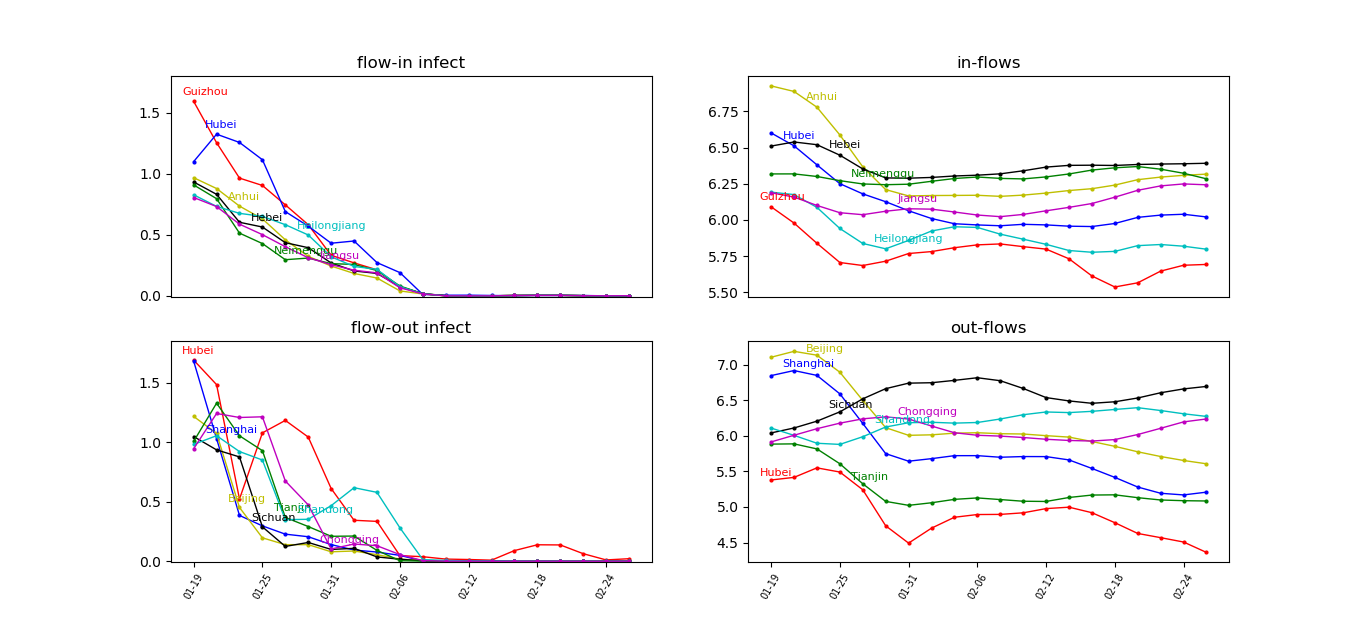}

     \caption{The temporal variation trend of the aggregated flow-in/-out infection link weight and population flow intensity}\label{co-variation}
\end{figure}

\subsection{Relaxation of population flows}

\begin{figure}[t!]

        \includegraphics[height=5.8cm,width=13cm]{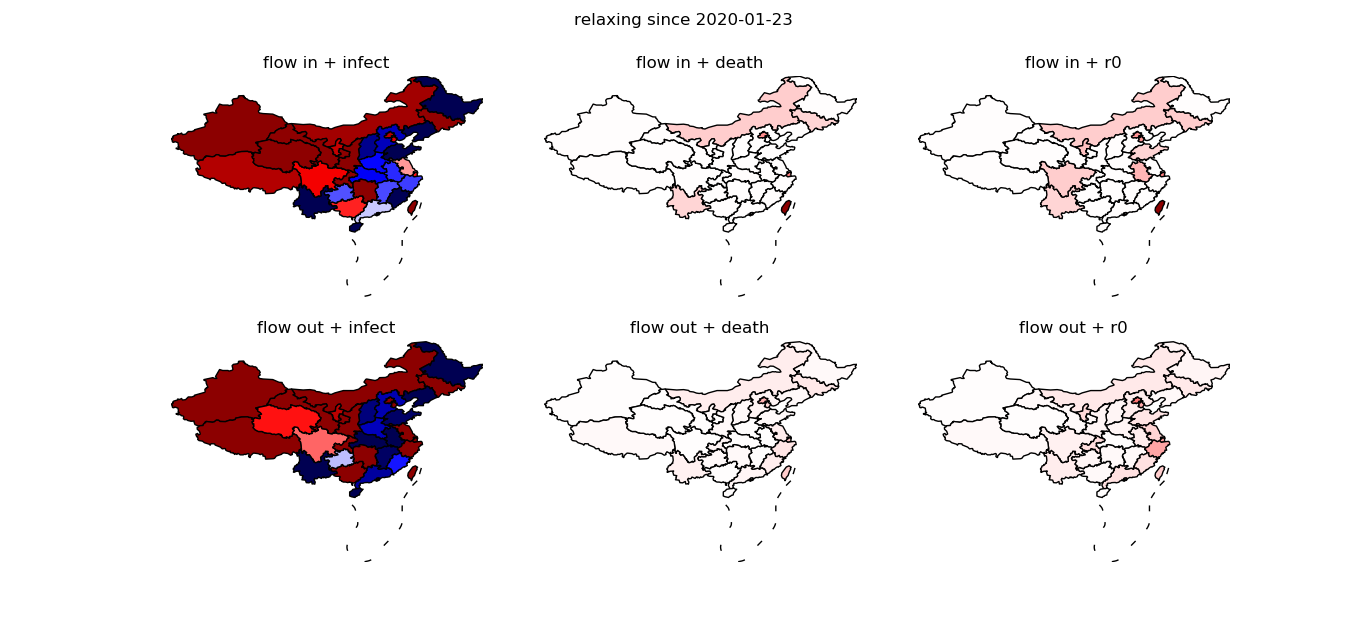}
        \includegraphics[height=5.8cm,width=13cm]{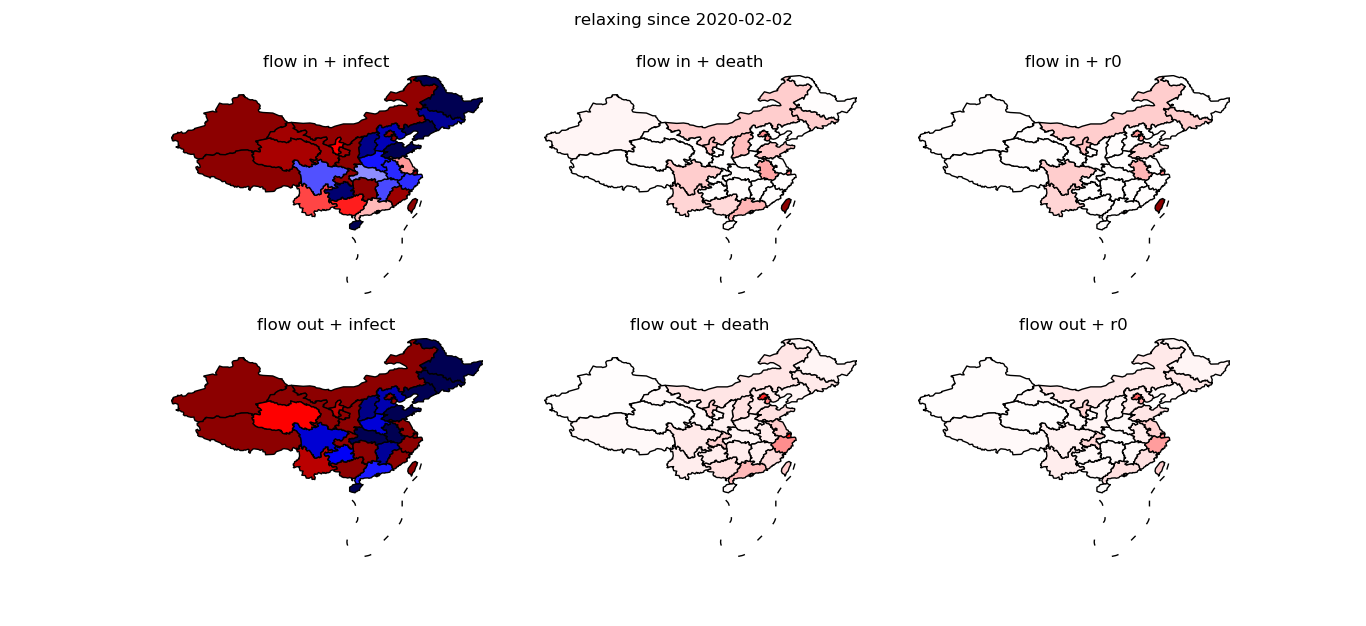}
        
 \includegraphics[height=7cm,width=13cm]{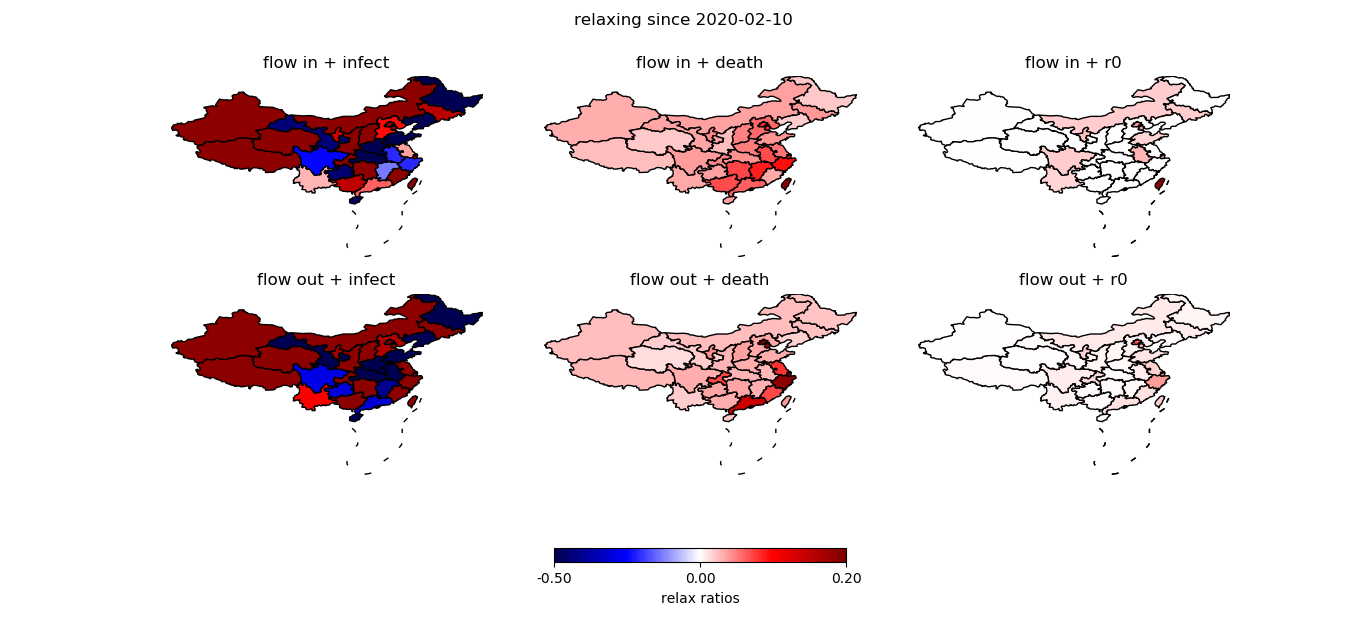}

     \caption{One set of optimal relaxation solutions to city-level travel ban since Jan. 23, Feb. 2 and Feb. 10, 2020}\label{relaxing map}
\end{figure}

As discussed in the previous section, the correlation between population flow intensity and infection link weight is weak in most cases. This observation implies that the most severe travel ban and lock-down may not be that necessary for those regions among which the infection connection is weak. Therefore, there should be potential to relax the lock-down even if the containment level of COVID-19 had to be maintained. To verify this argument, we solve the relaxation optimization problem stated in section \ref{counter}, the result is plotted within Fig. \ref{relaxing map}, where we plot the averaged ratio of the relaxation of population flow-in and flow-out intensity for all provinces in China. The ratio for every province $k$ (or $j$) is calculated through dividing the sum of all flow-in/-out index $\sum_{j}\mathbf{\overline{\mathbf{T}}}_{kj}(t_i)$ ($\sum_{k}\mathbf{\overline{\mathbf{T}}}_{kj}(t_i)$) by their optimally relaxed version $\sum_{j}\mathbf{\overline{\mathbf{T}}}^r_{kj}(t_i)$ ($\sum_{k}\mathbf{\overline{\mathbf{T}}}^r_{kj}(t_i)$), the average is taken over all $t_i$s after the relaxation starting date. Fig. \ref{relaxing map} displays the relaxation degree for all the three alternative starting date, Jan. 23, Feb. 02 and Feb. 10, 2020, and all the three containment targets \eqref{R0 compare}-\eqref{death compare}.

From Fig. \ref{relaxing map}, it is quite impressive that if control target is the total infectious cases, it seems impossible to relax the population flows between any pair of provinces without any more strict travel ban executed for Hubei and a couple of provinces that geographically connect to Hubei. And the impossibility of relaxation hold for all the three starting points. Such a result verifies the necessity of lock-down and strict travel ban executed by most of major cities in China since Jan. 23, 2020. This conclusion also agrees with the discussion in \cite{tian2020impact,li2020,qiu2020}. 

But on the other hand, if the target is to control the temporal $R_0$ that reflects the long-run infection severity and/or the total death number, the global relaxation becomes feasible even if starting from Jan. 23, 2020. In particular to the total death number, the travel restriction of all provinces in China can be relaxed substantially. For most of provinces in the south-eastern coast regions, the ratio of relaxation for flow-out population can exceed 10\%, while the flow-in relaxation ratio exceeds 5\%. In Zhejiang, Guangdong, Beijing, Shanghai and Tianjin, the flow-out ratio is even greater than 15\% and flow-in ratio is close to 10\%. As known, these five provinces consist of the most developed area of China in economy. A substantial relaxation of the traffic connection both within them and between them and the other provinces can significantly stimulate the overall economy growth for China. 

On the other hand, despite the existence of global relaxation strategy for the control target $R_0$, the potential for relaxation is not large. In the south-eastern coast area, most provinces have to maintain a strict travel ban at least in one direction (flow-in or out) in order to keep the $R_0$ reasonably low (in the sense of constraint condition \eqref{R0 compare}). This observation is partially because the index $R_0$ is much more sensitive, compared with the total death number, to the change of entries of $\mathbf{W}$, which restrict the space to relax the population flow. But compared with the total infection number, $R_0$ is less sensitive to the change of $\mathbf{W}$ because $R_0$ depends merely on the greatest positive eigenvalue of $\mathbf{W}$, while the infection number relies on every single entry. This explains why the global relaxation is still feasible for controlling $R_0$ but infeasible for controlling the total infections. 

Finally, if we come back to the target of controlling total infection, a partial relaxation strategy does exist after all (the partial relaxation arrangement is determined by maximizing the overall traffic flow intensity across all provinces, i.e. maximizing $\sum_{j,k,i}\mathbf{\overline{\mathbf{T}}}^r_{jk}(t_i)$, under the same set of constraints \eqref{R0 compare}-\eqref{death compare}, the overall traffic flow intensity can be viewed as an measure to the active degree of the economy). It is remarkable that since Feb. 02, 2020, if the travel ban was further strengthened for Hubei and its nearby provinces, the relaxation ratio for flow-out population becomes high for major south-eastern provinces, including Fujian, Zhejiang, Shanghai, Jiangsu, Beijing, Tianjin and Hebei, while the positive flow-in relaxation ratio is allowed to be positive for Guangdong. The existence of such an partial relaxation arrangement shows the existence of cross-regional substitutability of the strictness of lock-down, it also implies that a centralized decision mechanism for the choice of lock-down and travel ban could be more efficient in balancing the containment of COVID-19 outbreak and the economy resume.
%
%
%
%
%
%
%
%
%
%
%
%
%
%
%
%
%
%
%
%
%
%
%

In sum, by the counterfactual analysis on the relaxation of travel ban and city lock-down, we find that global relaxation strategies do exist for both the control target of $R_0$ and total death number, while it does not exist for control the total infection number, this observation results from the relative sensitivity between the control target variables and infection link weight $\mathbf{W}$. According to the degree of easiness in relaxing lock-down, controlling death is easier than controlling $R_0$, both of which are easier than controlling infection. To control the death number, a substantial relaxation has already been feasible since early Feb. 2020 for the major provinces in the south-east coast areas, relaxation for these provinces is critical to maintain the national-wide economy development. To control total infection, although a global relaxation is never feasible during the period studied in this paper, a partial relaxation is still possible by which the traffic intensity for south-eastern provinces can be relaxed substantially at the cost of a more strict lock-down for Hubei and the provinces that have close connection with Hubei. Such a partial relaxation arrangement is better for economy recovery, but its feasibility relies on the centralized decision on the lock-down as it does harm the local benefits via a more harsh travel ban.

\subsection{Relaxation of population flows under other non-lock-down measures}\label{alter measures}

In this section, we study the substitutability between lock-down and alternative non-lock-down measures. 
A further counter-factual analysis is carried out to reveal how the extent of population flow relaxation response to the strengthen of the non-lock-down measures.

\begin{figure}[t!]

        \includegraphics[height=8cm,width=13cm]{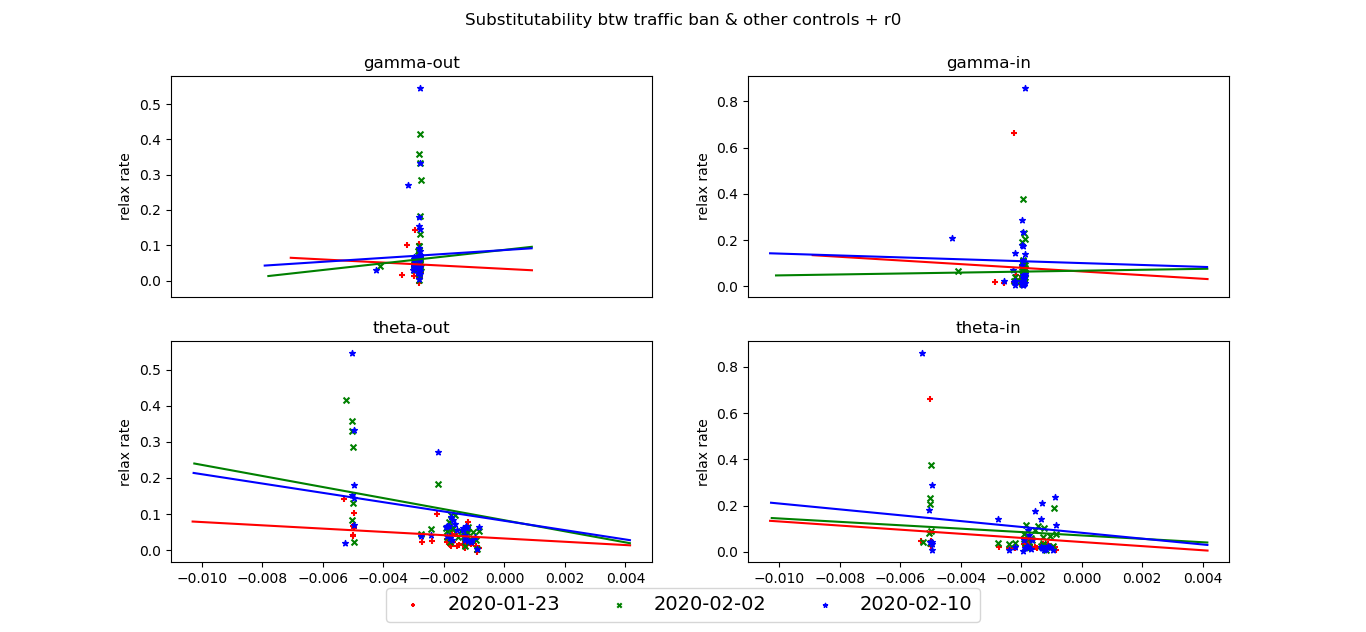}

     \caption{Substitutability between lock-down and non-lock-down measures given $R_0$ target}\label{r0 substitutability}
\end{figure}

\begin{figure}[t!]

        \includegraphics[height=8cm,width=13cm]{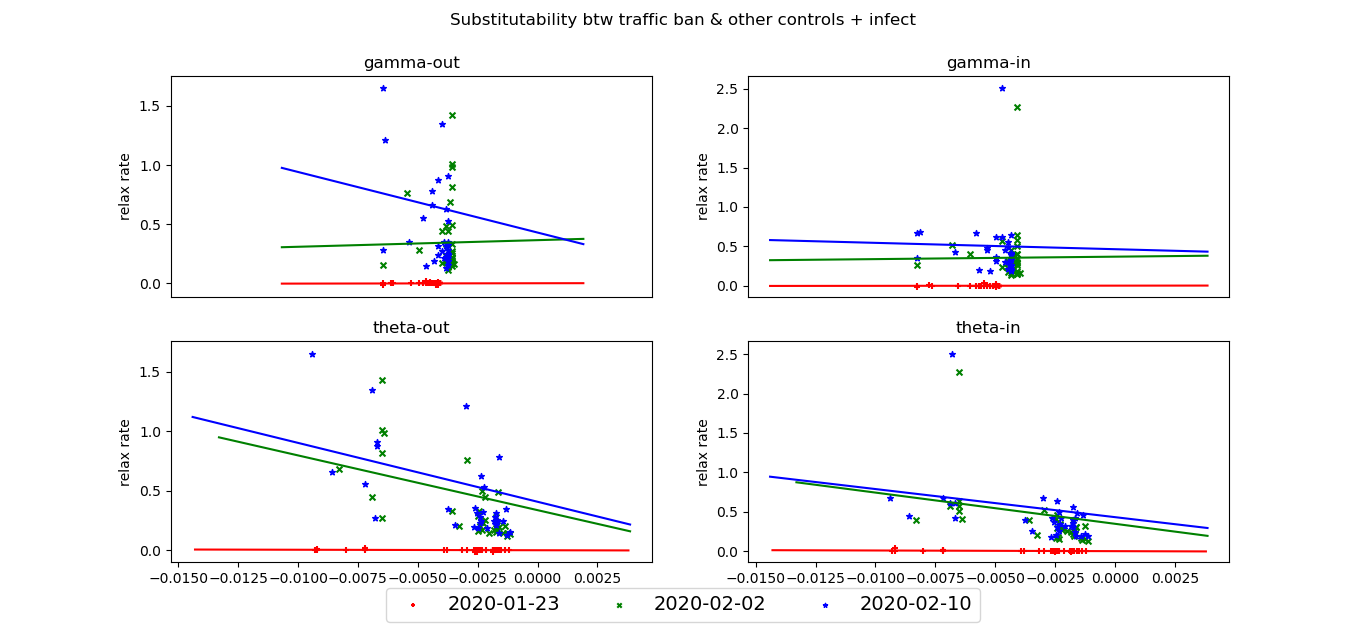}

     \caption{Substitutability between lock-down and non-lock-down measures under controlling the total infectious cases}\label{infect substitutability}
\end{figure}

\begin{figure}[t!]

        \includegraphics[height=8cm,width=13cm]{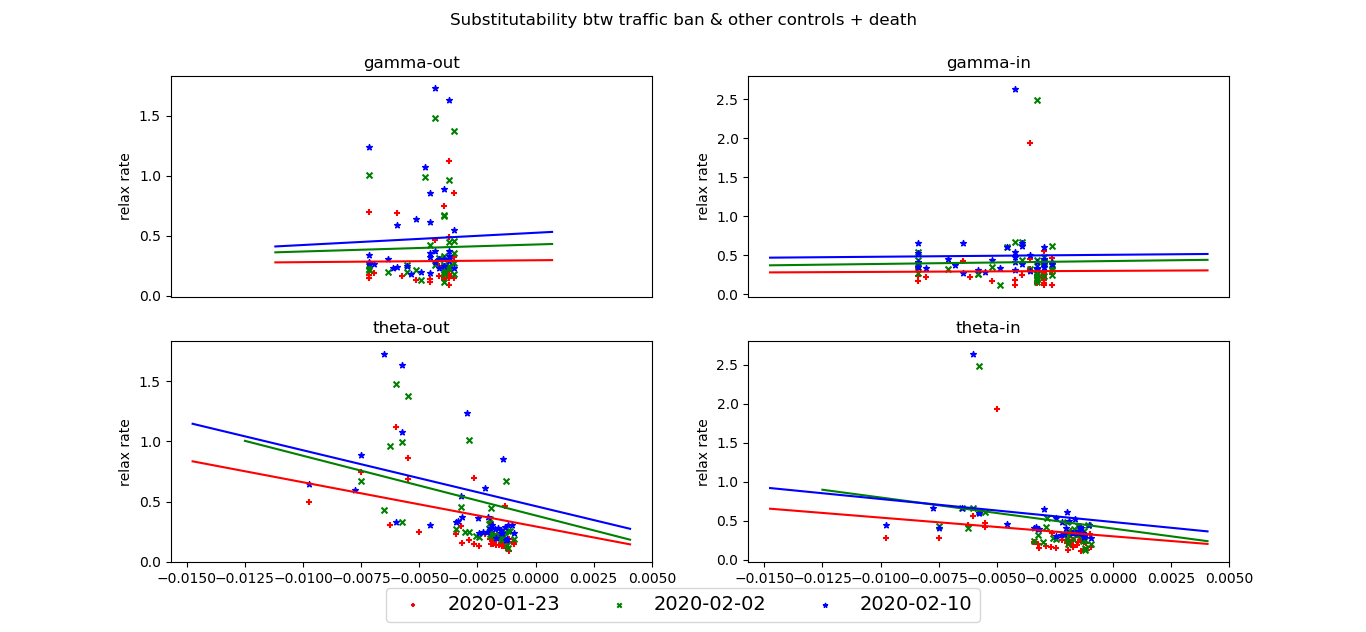}

     \caption{Substitutability between lock-down and non-lock-down measures under controlling the total dead cases}\label{death substitutability}
\end{figure}

Fig. \ref{r0 substitutability}-\ref{death substitutability} sketches the substitutability between the two classes of non-lock-down measures (their effect and executive strength are quantified by $\gamma_k$s and $\theta_j$s respectively) and the relaxation ratios of lock-down under three targets since three starting dates. As in the previous section, the relaxation ratios are aggregated according to the flow-in and -out direction on the province level and averaged over all time after the corresponding starting date. The first line subplots of Fig. \ref{r0 substitutability}-\ref{death substitutability} give the substitutability of the province-level $\gamma$s (horizontal axis) versus the flow-out(left)/flow-in(right) relaxation ratios (vertical axis); the second line presents the substitutability between the province-level $\theta$s (horizontal axis) and the the flow-out(left)/flow-in(right) relaxation ratios (vertical axis). The colored straight lines in each subplot correspond to the OLS-fitted line to the scatters with the same colors where the color is used to distinguish the three starting dates. From Fig. \ref{r0 substitutability}-\ref{death substitutability}, it is straightforward that there exists a gradually substitutable relationship between the non-lock-down measures by the flow-out region (represented by $\theta$s) and the relaxation ratios.  In addition, the substitutability between the $\theta$s and the flow-out relaxation ratios is stronger than between that and the flow-in relaxation ratios, this can be explained by that the $\theta$s is designed to capture the effect of such measures as setting health check-point in the high-way entrance, rail stations and airports. The main function of these measures is to reduce the potential infectious risk of flow-out population, therefore, they are more straightforwardly replacing the function of locking down all people within the city no matter whether they are healthy or not. In contrast, their effect on the flow-in relaxation ratios is via an indirect way. Compared to the substitutability of $\theta$s, there seems not to exist the gradual substitutability between the $\gamma$s and relaxation ratios. This is partially caused by the fact that the $\gamma$s represent the effect of the conditional quarantine measures executed by flow-in destinations and applied to suspected infectious cases and those travellers coming from out-town. These quarantine measures are not directly linked with the cross-regional population-flows and therefore affect the infection connection matrix $\mathbf{W}$ in an additive way. Compared to the multiplicative connection between $\theta$s and $\mathbf{W}$, the additive connection makes substitutability of $\gamma$s less direct. It is still impressive that most of the scatters in the second line subplots are clustered on the left of a vertical boundary line ($x\equiv c$ for some $c<0$) and a dense subset of these scatters are gathered around this boundary. In fact, this boundary phenomenon implies a much more stringent substitutability. That is, an universally bottom line exists such that the strength of flow-in non-lock-down measures cannot go below this line, otherwise it would squeeze the potential to relax the population flow intensity.

Through comparison across the target types and starting dates, it is found that for different targets, the degree of $\theta$'s substitutability is increasing in the order of controlling infection, $R_0$ and death. In particular, for the target of controlling infection number, there almost does not exist substitutability between $\theta$s and relaxation for the starting date Jan. 23, 2020 (reflected as the flat red line in the second line plots of Fig. \ref{infect substitutability}), which once again verifies the necessity of lock-down in the early stage. The order of substitutability is consistent with the order of easiness in relaxing population flow analyzed in the previous section, implying the relative easiness in the realization of different targets. For different starting date, the degree of substitutability of the $\theta$s and $\gamma$s is increasing for the later starting time, such as Feb. 2 and 10, 2020, which is reflected as (for $\theta$s) a greater absolute slopes of the green and blue lines than that of the red line in the second line plots of all three figures, and (for $\gamma$s) that the blue lines lie above green lines that lie above the red lines in the first line plots of Fig. \ref{infect substitutability} and \ref{death substitutability}. The increasing substitutability along with time support the story that the lock-down measure is effective in controlling the fast growth of infection number and the induced burden to the local healthcare system, which makes lock-down beneficial in the early stage of the explosion of community infection when there is no enough time left for figuring out all unknown infectious sources and no sufficient medical resources to conduct treatment. The lock-down in this stage can help save time for the effective reaction to the virus in the next stage and the adoption of more subtly designed prevention measures in the future. On the other hand, once if the explosion of community infection had been well contained and the total number of infectious cases were stablized, substitutability between lock-down and the other measures comes up, and it is proper to gradually turn lock-down to the other mild measures. Such a transition of containment measures agrees with the idea discussed in \cite{pueyo2020,harris2020}.

The next figure \ref{relaxing map1} presents the geographic distribution of relaxation ratios of flow-in and -out populations for different targets and different starting dates. The coloring scheme is exactly the same as that in Fig. \ref{relaxing map}. Comparing Fig. \ref{relaxing map1} with Fig. \ref{relaxing map}, it is quite surprising that for the starting date Feb. 2 and Feb. 10, 2020, almost all provinces (including Hubei province) in China can significantly relax their travel ban and lock-down policies, the relaxation ratios are almost uniformly greater than 20\% for both the flow-in and flow-out direction, and for both the targets of controlling total infection number and death number. For the target of $R_0$, the optimal relaxation ratios are a bit smaller than the other two targets, and the flow-out population flow of Hubei province cannot be relaxed even for the latest starting date Feb. 10, 2020. 

For the starting date Jan. 23, 2020, the optimal relaxation ratio does not change much compared to the later starting date for the target of controlling the death number of $R_0$, but a huge difference exist for the containment target of infection number.
If we counter-factually started the relaxation since Jan. 23, 2020, there is no global relaxation arrangement without increasing the infection number for some provinces and some time after Jan. 23, 2020. This conclusion is similar to that drawn from Fig. \ref{relaxing map}, it once again proves the robust necessity of lock-down in the early spreading stage of COVID-19.

It is remarkable to highlight the difference in the absolute size of relaxation ratios between the existence and non-existence of adjustment
 to the stringency of non-lock-down measures since Feb. 2, 2020. 
In the later situation, the value of relaxation ratios is almost uniformly twice greater that that in the former situation. This fact implies the existence of a better combination of various control measures during the China's anti-COVID-19 movement. That is the execution of lock-down for a very short period since Jan. 23, 2020 (e.g. one week) in order to save time for stablizing the infection number and meanwhile preparing for the transition to the other milder measures, such as the conditional quarantine and health check-points. Then gradually relax the degree of lock-down since Feb. 02, 2020 through substituting with an increasingly stringent execution of the other non-lock-down measures. Such a quick lock-down strategy, compared to the 1-month+ lock-down that was actually carried out in the real time line, have the least harm to the economy while can reach the same effect on mitigating the outbreak of COVID-19.

%
%


\begin{figure}[t!]

        \includegraphics[height=5.8cm,width=13cm]{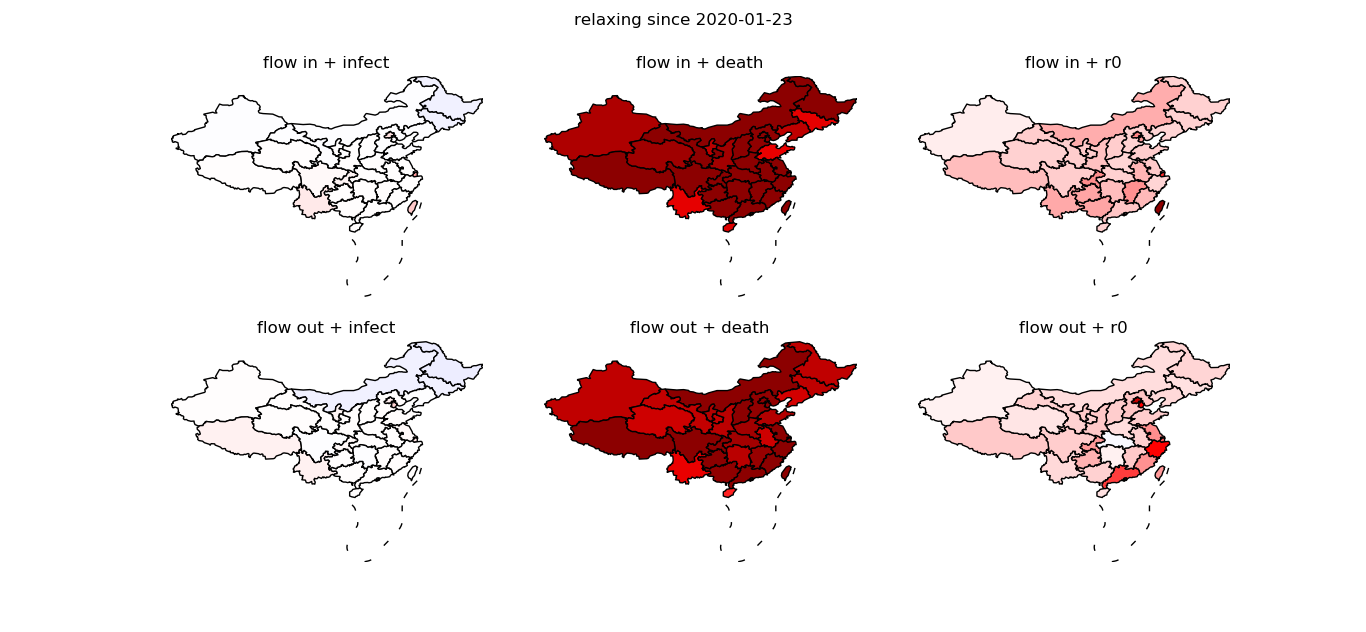}
        \includegraphics[height=5.8cm,width=13cm]{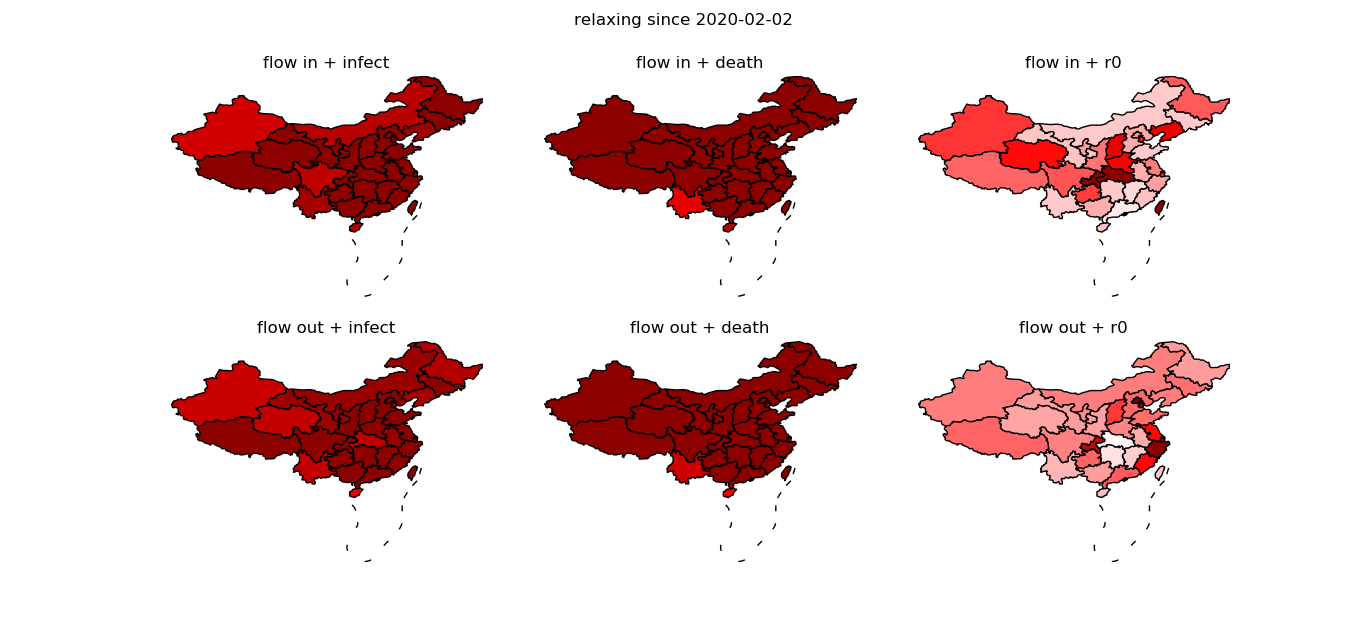}
        
 \includegraphics[height=7cm,width=13cm]{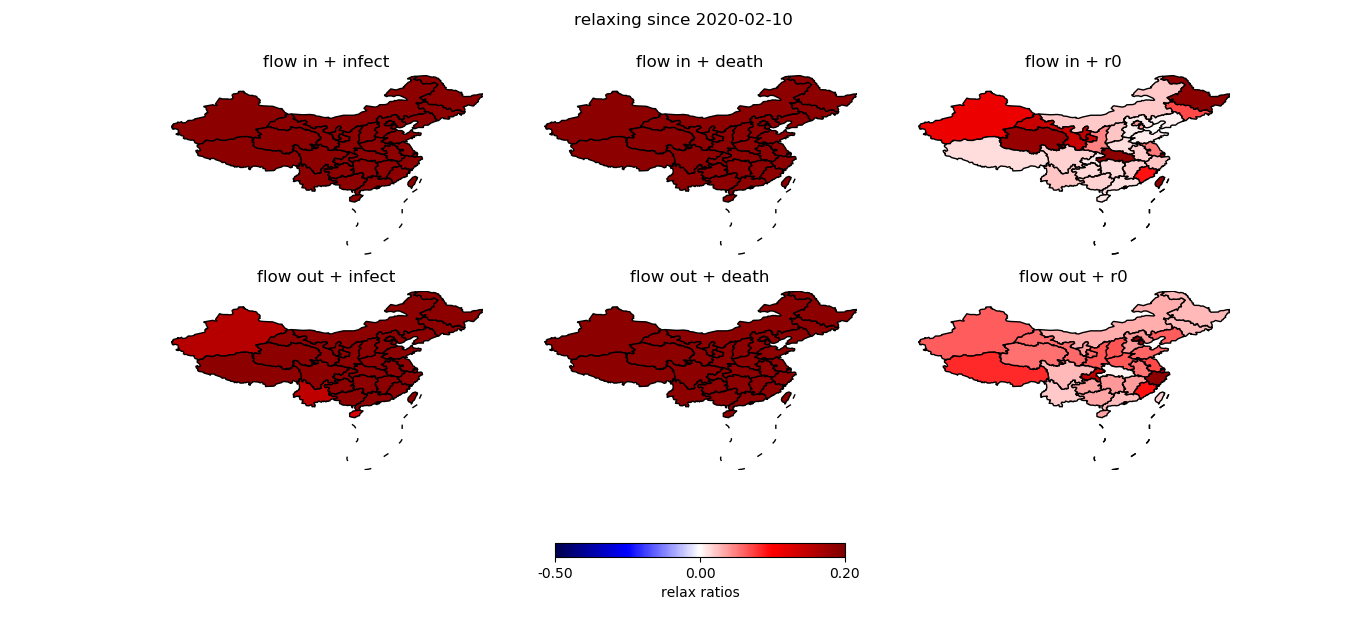}

     \caption{One set of optimal relaxation solutions to lock-down since Jan. 23, Feb. 2 and Feb. 10, 2020 under the existence of adjustable non-lock-down measures}\label{relaxing map1}
\end{figure}

\section{Discussion \& Conclusion}

In this study, we propose a non-parametric network-based SIR model (NP-Net-SIR) to study the cross-regional outbreak of COVID-19, within which the special epidemic characteristics of COVID-19, such as the long incubation period and asymptomatic infection channel, are easily encoded. The non-parametric nature of NP-Net-SIR saves it from suffering the presumed liner dependence between COVID-19 outbreak and the inter-regional population flow, which might lead to over-estimate of the real effect of city lock-down. The low accuracy of outbreak data before the mid of Feb. 2020 imposes a major technical challenge to those studies based on statistic inference from the early outbreak. To resolve the data issue, we apply the graph-Laplacian regularization from semi-supervised learning to identify and train NP-Net-SIR model which turns out robust under poor data quality condition. 

By the trained model, we analyze the connection between population flow and the cross-regional infection network, based on which a set of counter-factual analysis is carried out to study the necessity of lock-down and substitutability between lock-down and the other containment measures. The main findings of this study include: 1) except for the very early stage of outbreak and the population flow out of the epicenter Wuhan and Hubei province, there does not exist strong linear connection between population flow and cross-regional infection connection, indicating that the lock-down may not be the key measure to contain the COVID-19; 2) strong substitutability exists between the lock-down and non-lock-down-typed containment measures, between different containment targets, and between the lock-down of different regions; 3) in the earliest stage (starting from Jan. 23, 2020) the lock-down of the epicenter, Hubei, is indispensable, while the indispensability is by and large attributed to the geographically unbalanced impact of the COVID-19 outbreak and the cross-regional inequality in terms of the public awareness of the COVID-19, healthcare resources and the implementation of containment measures; 4) after the impact of COVID-19 got equalized inter-regionally (e.g. after Feb. 2, 2020), the lock-down had already been able to be relaxed substantially while the same containment effect can be achieved; 5) when the other containment measures are implemented stringently, the relaxation degree of population flow can be even enlarged. 

Our findings support that the lock-down may not be the optimal strategy in containing the outbreak of COVID-19 except for the early stage, there exist alternatives that have less negative impact on the social-economic development. But the effectiveness of the alternative measures requires a subtly designed prevention system which should admit the regional difference and the temporal adjustment in the containment measures according to the particular situations for different regions and different time periods. The discussion in this paper has certain guiding and practical significance for the normalization of the epidemic prevention, the resumption of production and economic activities from lock-down, and the containment strategy design of other countries in the same epidemic situation.

Although the analysis of this paper is retrospective and based on that all the data of COVID-19 have been available, which is not possible for the decision time at Jan. 23 and the early Feb., 2020, it is still meaningful to retrospect the potential optimal controlling strategy. This is because even by now, China is still facing a high risk of the \enquote{second-wave} outbreak. The choice of both feasible and effective containment
 measures is still a critical but open question, while many countries in the world currently still struggle with how to prevent the outbreak of COVID-19. Our study can provide some hints on this choice. First, the China's experience and the strict lock-down measure turns out not only sufficient \citep{tian2020impact,li2020,qiu2020,fang2020,prem2020,tuite2020} for mitigating the virus spread, but may also be the only effective way to cool down the explosion of community infection at least in the early spreading stage. But afterward, it shouldn't be stuck in the lock-down status for long which is neither meaningful for control the virus nor good for the economic recovery. In contrast, a set of non-lock-down-typed alternative measures should be quickly prepared and actively executed so as to substitute the lock-down which, as long as being strictly executed, can lead to as effective control of the virus as the lock-down can do. Meanwhile, without the collaboration of the non-lock-down-typed measures, such as the conditional quarantine, the purely lock-down may also fail to mitigate the COVID-19, as what happened in Italy, Spain, and New York, USA.  
\section*{Acknowledgement:}
\noindent This work was partially supported by the Ministry of Education in China Project of Humanities and Social Sciences under Grant No. 20YJC790176, and the Fundamental Research Funds for the Central Universities under Grant No. 2242020S30024.

\bibliographystyle{apalike}

\begin{appendix}
\section{Training of NP-Net-SIR model}\label{training}
The non-parametric network set-up makes our NP-Net-SIR model essentially a special class of recurrent neural network (RNN), namely the temporal RNN \citep{liu2016} with the temporality coming from the time dependent neural network $\mathbf{W}(t)$. The total amount of infections $\mathbf{n}(t)$, due to its unobservable nature, corresponds to the hidden layer of the RNN, while the documented infections $\mathbf{m}(t)$ corresponds to the output layer. Due to the lack of extra input to the NP-Net-SIR, the input layer is degenerated to $0$. Given the observation of the sequence of documented infections $\mathcal{M}_o=\{\mathbf{M}_{t_i}:\,i=1,\dots,n;\,t_1<\cdots<t_n\}$ and a proper regularized loss function, the standard back-propagation method applies to estimate the set of unknown temporal parameters $\{\mathbf{W}(t_i),\mathbf{n}(t_i),r(t_i),p_I(j,t_i),p_B(j,t_i):\,i=1,\dots,n;\,j=1,\dots,incub\}$. Due to the discreteness of the observation time, the continuity condition for these temporal parameters can be converted to a graph-Laplacian regularization with the grid graph on real line \citep{zhou2011}, which is asymptotically equivalent to require, under the high-frequent observation, these temporal parameters are continuous, differentiable and have square-integrable derivatives. In our special case, the graph-Laplacian regularization can be written in the following form:
\begin{equation}\label{regularization}
\begin{aligned}
\mathcal{R}(\mathbf{W}, r, p_B, p_I)=&\Vert \mathbf{W}(t_n)\Vert+\sum_{i=1}^{n-1}\Vert \mathbf{W}(t_{i+1})-\mathbf{W}(t_i)\Vert^2+\sum_{i=1}^{n-1}\Vert r(t_{i+1})-r(t_i)\Vert^2+\\
&\sum_{i=1}^{n-1}\sum_{j=1}^{incub}\left(\Vert p_B(j,t_{i+1})-p_B(j,t_i)\Vert^2+\Vert p_I(j,t_{i+1})-p_I(j,t_i)\Vert^2\right)
\end{aligned}
\end{equation}
where we artificially set the boundary $\mathbf{W}(t_{n+1})\equiv 0$ so that the summation of hidden networks $\mathbf{W}$ up to the supscript $n$ implies the sparse requirement on the $\mathbf{W}(t_i)$'s which is standard to avoid over-fitting. 

For loss function, in addition to the standard square-sum error between the observed $\mathbf{M}_{t_i}$s and the estimated $\mathbf{m}(t_i)$s, we add an extra penalty to the error function in order to fix the data pollution issue in the early stage of COVID-19 outbreak. In particular, we define the following indicator function:
\begin{equation}\label{bias}
I_{t^\ast}(t,m,M)=\begin{cases}
m-M & if \, t\geq t^\ast\textrm{ or } m>M\\
0 & else
\end{cases},
\end{equation}
the meaning of \eqref{bias} is that there is a cut-off time point $t^\ast$ before which the documented infection number tends to under-estimate the real spreading trend. Therefore, if the estimated number $m$ exceeds the reported $M$ we think the estimates reflect the true case and don't treat it as an error, while if the estimated is less than the reported, which indicates a severe under-estimate to the true case, the error is calculated as usual. After $t^\ast$, it is thought that all hidden infectious cases that should be documented and published have already been reported, then the reported cases agree with the real trend. In this paper, we set $t^\ast$ as the date, Feb. 12, 2020, when Wuhan local government reported 13,000+ inventory infectious cases that were not in record before. Then the loss function can be written as the following form:
\begin{equation}\label{loss}
L(\mathcal{M}_o,\mathbf{m},\mathbf{n},\mathbf{W}, r, p_B, p_I)=\sum_{i=1}^{n}\sum_{j=1}^{k}\Vert(M_{t_i,j}-m_j(t_i))*I_{t^\ast}(t_i,m_j(t_i),M_{t_i,j})\Vert^2+\mathcal{R}(\mathbf{W}, r, p_B, p_I),
\end{equation} 
where the loss depends on the hidden infection number $\mathbf{n}$ through the observed infection number $\mathbf{m}$ via model \eqref{network sir}.

Note that the RNN nature of the model \eqref{network sir} makes the infection number $\mathbf{n}(t)$, $\mathbf{m}(t)$ generated from $\mathbf{n}(s)$, $\mathbf{m}(s)$ for $s<t$, then by the back-propagation algorithm, the model \eqref{network sir} is fitted in a reversed order, i.e. the parameter values of $\mathbf{n}(s)$, $\mathbf{W}(s)$, $p_I(\cdot,s)$, $p_B(\cdot,s)$ and $r(s)$ for previous period $s$ are essentially fitted from the later observations $\mathbf{m}(t)$ with $t>s$. The back-ward fitting direction together with the function \eqref{bias} presents a way to utilize the label data $\mathbf{m}(t)$ at time $t>t^\ast$ to generate label of infection number for those un-labeled time $s$ with $s\leq t^\ast$, such a trick of utilizing partially labeled data is standard in semi-supervised learning \citep{zhou2011}, we borrow it here to address the inaccurate data issue for the early stage.
  
To estimate the parameters, we minimize the loss function with respect to parameters and also subject to the default range restrictions that are the following:
\begin{equation}\label{cons}
\begin{cases}
0\leq \mathbf{W}_{kl}(t_i)\leq 1,& \forall k,l,i\\
r(t_i)\geq 0,&\forall i\\
p_B(j,t_i),p_I(j,t_i)\geq 0,&\forall j,i\\
\sum_{j}p_B(j,t_i)=\sum_{j}p_B(j,t_i)=1,&\forall i
\end{cases}
\end{equation} 
The quadratic nature of the square-sum loss function guarantees that even if the penalty \eqref{bias} is added, the resulting loss function \eqref{loss} is still continuously differentiable, standard gradient descending solvers are applicable. 

\section{Training algorithm}\label{train}
Training model \eqref{network sir} is equivalent to solving the optimization problem in \eqref{loss} under the constraints \eqref{cons}. The classical gradient-descending-based solution for RNN usually assumes no constraint. Therefore, some modifications are needed. In the following, we propose a sequential modification to the classical backward propagation training algorithm for neural network model. To facilitate the introduction of the sequential algorithm, we temporally assume the temporal RNN is no longer temporal, but a static RNN, i.e. all the temporal parameters $\mathbf{W}$, $p_B$, $p_I$ and  $r$ are no longer dependent on $t$. Also suppose that the infection number $\mathbf{m}_t$ is observed within the discrete time interval $\{1,\dots,T\}$. Then, the discrete version of model \eqref{network sir} under above assumptions becomes the following: 
\begin{equation}\label{mean-field equation}
\begin{aligned}
\Delta \mathbf{n}_{t}&=\mathbf{n}_{t+1}-\mathbf{n}_{t}=\sum_{i=1}^{incub}p_{I,i} \mathbf{W}\cdot \mathbf{n}_{t-i}-r\mathbf{n}_t\\
\mathbf{m}_{t}&=\sum_{i=1}^{incub}p_{B,i}\cdot \mathbf{n}_{t-i}
\end{aligned},
\end{equation} 
where for $x=B,I$, $p_{x,i}$ is a short-hand notation for $p_x(i)$ when the static probability $p_x$ mass function is evaluated at the discrete time $i$. Given \eqref{mean-field equation}, notice that when the unknown model probability parameter $p_{B}$, $p_{I}$, the recovery rate $r$ and network matrix $\mathbf{W}$ are fixed, the model depends completely on the hidden layer $\mathbf{n}=\{\mathbf{n}_t:\,t=-incub,-incub+1,\dots,0,\dots,T\}$ via vector multiplication. While $p_{B}$, $p_{I}$, $r$ and $\mathbf{n}$ are fixed, the model depends completely on $\mathbf{W}$ via matrix multiplication. When $\mathbf{W}$ and $\mathbf{n}$ are fixed, the model depends completely on the $p_{B}$, $p_{I}$ and $r$ via constant multiple and vector inner product. Note that all above operations are linear operations, meanwhile, the loss function \eqref{loss} has quadratic form, these facts imply that fixing any two classes of quantities among (a) $p_{B}$, $p_{I}$, $r$; (b) $\mathbf{W}$; and (c) $\mathbf{n}$, the optimization problem \eqref{loss} under constraint \eqref{cons} is a classical convex programming problem \citep{shen2014reconstructing,Golstein2008}, with respect to the remaining class of quantities. As our loss function \eqref{loss} is strictly convex, the 
resulting convex programming problem has the unique minimum and can be solved quickly via the classical gradient algorithm. Therefore, under static setting of model parameters, the following iterative fitting algorithm can be applied to train the parameters:\\
\noindent {\bf Step 1:} Given $s\geq 0$, for fixed vector $p^s_{B}$, $p^s_{I}$, constant $r^s$ and matrix $\mathbf{W}^s$, solve problem \eqref{loss} under \eqref{cons} with respect to $\mathbf{n}$, resulting in $\mathbf{n}^{s+1}$;\\

\noindent {\bf Step 2:} Given $s\geq 0$, for fixed vector $p^s_{B}$, $p^s_{I}$, constant $r^s$ and time series $\mathbf{n}^s$, solve problem \eqref{loss} under \eqref{cons} with respect to $\mathbf{W}$, resulting in $\mathbf{W}^{s+1}$;\\

\noindent {\bf Step 3:} Given $s\geq 0$, for fixed matrix $\mathbf{W}^s$ and time series $\mathbf{n}^s$, solve problem \eqref{loss} under \eqref{cons} with respect to vector $p_{B}$, $p_{I}$ and constant $r$, resulting in $p^{s+1}_{B}$, $p^{s+1}_{I}$ and $r^{s+1}$;\\

\noindent {\bf Step 4:} Repeat {\bf Step 1-3} until the ratio of $L^2$ norms \begin{equation}
    \frac{\Vert p^{s+1}_{B}-p^{s}_{B} \Vert+\Vert p^{s+1}_{I}-p^{s}_{I} \Vert+\Vert r^{s+1}-r^{s} \Vert+\Vert\mathbf{W}^{s+1}-\mathbf{W}^{s} \Vert}{\Vert p^{s}_{B} \Vert+\Vert p^{s}_{I} \Vert+\Vert r^{s} \Vert+\Vert\mathbf{W}^{s}\Vert}
\end{equation}
is less than a prescribed threshold $\delta$ (=$10^{-3}$).

Then, to release the static assumption, given the data $\mathcal{M}_o=\{M_0,M_1,\dots,M_S\}$ of the series of observed infection vector during the period end up with day $S$. consider the following sequential backward propagation\\

\noindent {\bf Step 1:} (Initialization) Set $\tau=S$ $\mathcal{M}_\tau=\{M_{\tau-T},\dots,M_\tau\}$, apply the 4-step static training algorithm as above, denote the output as $\mathbf{W}^\tau$, $p_{B}^\tau$, $p_{I}^\tau$, $r^\tau$ and $\mathbf{n}^\tau=\{n^\tau_{\tau-T-incub},\dots,\mathbf{n}^\tau_{\tau}\}$;\\
\noindent {\bf Step 2:} For $T\leq\tau<S$ and $\mathcal{M}_\tau$, redefine the hidden vector as $\mathbf{n}=\{\mathbf{n}_{\tau-T-incub},\mathbf{n}^{\tau+1}_{\tau-T-incub+1},\\
\dots,\mathbf{n}^{\tau+1}_\tau\}$ where only the first entry $\mathbf{n}_{\tau-T-incub}$ is undetermined and needs to be optimized, the remaining entries are fixed via the estimated value from the previous iteration. Given the estimation $\mathbf{W}^{\tau+1}$, $p_{B}^{\tau+1}$, $p_{I}^{\tau+1}$, $r^{\tau+1}$ from the previous iteration, apply the static version of training algorithm as above with the redefined loss function as in the following equation \eqref{loss1}, we get the output $\mathbf{W}^\tau$, $p_{B}^\tau$, $p_{I}^\tau$, $r^\tau$ and $\mathbf{n}^\tau=\{n^\tau_{\tau-T-incub},\mathbf{n}^{\tau+1}_{\tau-T-incub+1},\dots,\mathbf{n}^\tau_{\tau}\}$.
\begin{equation}\label{loss1}
\begin{aligned}
L(\mathcal{M}_\tau,\mathbf{m},\mathbf{n}_{\tau-T-incub},\mathbf{W},r,p_B,p_I,)=&\sum_{i=\tau-T+1}^{\tau}\sum_{j=1}^{k}\Vert(M_{i,j}-\mathbf{m}_{i,j})*I_{t^\ast}(i,\mathbf{m}_{i,j},M_{i,j})\Vert^2+
\Vert \mathbf{W}^{\tau+1}-\mathbf{W}\Vert^2\\
&+\Vert r^{\tau+1}-r\Vert^2+
\Vert p^{\tau+1}_B-p_B\Vert^2+\Vert p_I^{\tau+1}-p_I\Vert^2
\end{aligned}
\end{equation} 
where the timely integrated Laplacian regularization \eqref{regularization} in loss function \eqref{loss} is replaced with the one-period regularization. 

Combining the sequence of outputs from the two-step sequential backward propagation algorithm, we obtain the estimated sequence of adjacency matrices $\{\mathbf{W}^T,\dots,\mathbf{W}^S \}$, probability parameters $\{p_B^T,\dots,p_{B}^S\}$, $\{p_I^T,\dots,p_{I}^S\}$, recovery rate $\{r^T,\dots,r^S\}$ and the sequence of hidden infection vector $\{\mathbf{n}_{-incub},\mathbf{n}_{-incub+1},\dots,\mathbf{n}_{0},\dots,\mathbf{n}_S\}$. For the hidden infection vector, note that the estimate to $\mathbf{n}_\tau$ for every $\tau$ is unique according to the design of $\mathbf{n}^\tau$ in the {\bf Step 2} of the sequential backward propagation.

The sequential backward propagation is essentially a sequence of the standard backward propagation which is applied to solve the static version of our model \eqref{mean-field equation}, where the connection between two consecutive steps is established through the consecutive one-period decomposition of the Laplacian regularization condition in \eqref{loss1} and the construction that let $\mathbf{n}_\tau$ and $\mathbf{n}_{\tau+1}$ share the common hidden infection numbers in the overlapped period. It is not hard to verify that the sequential implementation of backward propagation generates asymptotically equivalent result to the classical backward propagation.

Also notice that the sequential training depends on an unspecified horizon parameter $T$, in this paper, we set $T=7$ as it minimizes the aggregated loss $\eqref{loss}$ compared to the alternatives in the range $\{1,\dots,20\}$. The implementation of the algorithm is by python where the key-step minimization ({\bf Step 1-3}) for the static model \eqref{mean-field equation} is implemented via the convex programming package, CVXPY.

\section{Calculation death number}\label{death calc}
$\mathbf{D}(t_i)$ (similarly $\mathbf{D}^r(t_i)$)
is calculated from the sequence $\{\mathbf{m}(t):\,t\leq t_i\}$ through following auto-regressive equation
\begin{equation}\label{reg2}
\mathbf{Dr}_j(t_i)=a_j+b\cdot \Delta\tilde{m}_{j}(t_{i-k_1})+c\cdot \mathbf{Dr}_j(t_{i-k_2})+d\cdot h_j(t_i)+\varepsilon_{j}(t_i)
\end{equation}
where $\mathbf{Dr}_j(t_i)$ is the death rate such that $\mathbf{D}_j(t_i)=\mathbf{Dr}_j(t_i)\mathbf{m}_j(t_i)$. In \eqref{reg2} $h_j(t_i)$ is the ratio between $\mathbf{m}_j(t_i)$ and the local healthcare resources that are measured by the total number of hospital beds. According to preliminary analysis, the time lag $k_1$, $k_2$ take 8 day and 1 day, respectively. The coefficient and residuals in \eqref{reg2} are inferred from the real data, they will be fixed in the counterfactual analysis and help generate the updated death number for counterfactually adjusted input.

\end{appendix}
\end{document}